\documentclass[fleqn,usenatbib,useAMS]{mnras}

\usepackage{lmodern}
\usepackage[T1]{fontenc}
\usepackage{amsmath}
\usepackage{amssymb}
\usepackage[a4paper]{geometry}
\usepackage{hyperref}

\usepackage{url}
\usepackage{multirow}
\usepackage{caption}
\usepackage[labelformat=simple]{subcaption}

\usepackage[nodayofweek]{datetime}
\usepackage{enumerate}
\usepackage{xspace}
\usepackage{xcolor}
\usepackage{graphicx}

\usepackage{import}
\usepackage{tabularx, booktabs}
\newcolumntype{Y}{>{\centering\arraybackslash}X}

\usepackage{siunitx}
\usepackage{lipsum}
\usepackage{mathtools, cuted}

\graphicspath{{figures/}}

\usepackage{graphicx}	
\usepackage{amsmath}	
\usepackage{multicol}        
\usepackage{bm}		
\usepackage{pdflscape}	
\usepackage{lmodern}
\usepackage{fix-cm}
\usepackage[T1]{fontenc}
\usepackage{ae,aecompl}

\usepackage{newtxtext,newtxmath}
\usepackage{wrapfig}

\newcommand{\gaia}{\emph{Gaia}\xspace}
\newcommand{\skyc}{\mbox{Sky}\textsc{CURTAIN}s\xspace}
\newcommand{\SBone}{SB1\xspace}
\newcommand{\SBtwo}{SB2\xspace}

\newcommand{\FfF}{\textsc{Curtain}sF4F\xspace}
\newcommand{\flowsforflows}{Flows for Flows\xspace}
\newcommand{\cwola}{{CWoLa}\xspace}
\newcommand{\vm}{\textsc{Via Machinae}\xspace}

\newcommand{\gdone}{\mbox{GD-1} stream\xspace}

\newcommand{\lon}{\ensuremath{\phi}\xspace}
\newcommand{\lat}{\ensuremath{\lambda}\xspace}
\newcommand{\ra}{\ensuremath{\alpha}\xspace}
\newcommand{\dec}{\ensuremath{\delta}\xspace}
\newcommand{\pmra}{\ensuremath{\mu_{\alpha}}\xspace}

\newcommand{\pmracosdec}{\ensuremath{\mu_{\alpha} \cos \delta}\xspace}
\newcommand{\pmdec}{\ensuremath{\mu_{\delta}}\xspace}
\newcommand{\pmlon}{\ensuremath{\mu_{\lon}}\xspace}
\newcommand{\pmlonstar}{\ensuremath{\mu_{\lon}^*}\xspace}

\newcommand{\pmloncoslat}{\ensuremath{\mu_{\lon} \cos \lat}\xspace}
\newcommand{\pmlat}{\ensuremath{\mu_{\lat}}\xspace}
\newcommand{\parallax}{\ensuremath{\varpi}\xspace}
\newcommand{\magnitude}{\ensuremath{G}\xspace}
\newcommand{\gbprp}{\ensuremath{G_{\text{BP}}-G_{\text{RP}}}\xspace}

\newcommand{\nvidia}{\mbox{NVIDIA\textsuperscript{\textregistered}}\xspace}

\title[\skyc]{\skyc: Model agnostic search for Stellar Streams with Gaia data}

\author[D. Sengupta et al.]{
\thanks{Contact e-mail: \href{mailto:debajyoti.sengupta@unige.ch}{debajyoti.sengupta@unige.ch}}%
Debajyoti Sengupta,$^{1}$
Stephen Mulligan,$^{1}$
David Shih,$^{2}$
John Andrew Raine,$^{1}$
and Tobias Golling$^{1}$
\\
$^{1}$Département de physique nucléaire et corpusculaire, University of Geneva, Switzerland\\
$^{2}$NHETC, Dept. of Physics and Astronomy, Rutgers, Piscataway, NJ 08854, USA}

\date{}

\pubyear{2024}

\begin{document}

\maketitle

\begin{abstract}
We present \skyc, a data driven and model agnostic method to search for stellar streams in the Milky Way galaxy using data from the \gaia telescope.
\skyc is a weakly supervised machine learning algorithm that builds a background enriched template in the signal region by leveraging the correlation of the source's characterising features with their proper motion in the sky.
This allows for a more representative template of the background in the signal region, and reduces the false positives in the search for stellar streams.
The minimal model assumptions in the \skyc method allow for a flexible and efficient search for various kinds of anomalies such as streams, globular clusters, or dwarf galaxies directly from the data. 
We test the performance of \skyc on the \gdone and show that it is able to recover the stream with a purity of 75.4\% which is an improvement of over 10\% over existing machine learning based methods while retaining a signal efficiency of 37.9\%.
\end{abstract}

\begin{keywords}
Galaxy: Stellar Content -- Galaxy: Structure -- Stars: Kinematics and Dynamics -- Methods: Weakly Supervised Machine Learning
\end{keywords}

\section{Introduction}
\label{sec:intro}

When smaller gravitationally bound systems such as globular clusters or satellite dwarf galaxies, are disrupted by their host galaxy, the stars in these systems are tidally stripped off.
This results in a stream of stars, named \textit{stellar streams}, which, over time trace out the orbit of the progenitor system.
Since the interactions between these large-scale gravitationally bound systems occur over a very long timescale, real time observations of these events are impossible.
Stellar streams are therefore an excellent alternative probe into the merger history of these systems~\citep{stream_merger1,stream_merger2,stream_merger3,stream_merger4,stream_merger5}.
Moreover, the orbits of these streams are sensitive to the gravitational potential of the host galaxy, and thus can be used to constrain the mass distribution in it~\citep{stream_potential1,stream_potential2,stream_potential3,stream_potential4,stream_potential5}.
Over time, due to gravitational interaction with the surrounding matter, the shape of these streams change, and the density perturbations therein, such as gaps and spurs, can also provide insights into the dark matter distribution in the galaxy~\citep{stream_dark0,stream_dark1,stream_dark2,stream_dark3,stream_dark4} and its properties~\citep{stream_darkprop1,stream_darkprop2}.
The study of stellar streams is thus crucial to understanding the formation and evolution of galaxies, and the content thereof.

The \gaia mission~\citep{gdr2} has provided an unprecedented dataset of stars in the Milky Way, with accurate astrometric and photometric measurements.
This wealth of data has allowed the development of several techniques to detect stellar streams~\citep{streamfinder_18a,streamfinder_18b,stargo_18,stream_find1,stream_find2,stream_find3,stream_find4}.
In general, these methods leverage the astrophysics of stellar streams, such as their grouping in chemical composition and kinematics, to identify the stream candidates.
For instance, the \textsc{Streamfinder} algorithm~\citep{streamfinder_18a,streamfinder_18b} assumes a specific model for the gravitational potential of the Milky Way galaxy, and searches for stars occupying the same hyperdimensional tubes through a six-dimensional positional and velocity space.

More recently, several machine learning techniques have been employed to detect stellar streams.
Particularly, \vm~\citep{vm1,vm2}, and \cwola~\citep{astrocwola} are fully data-driven and have very minimal model assumptions about the streams.
These techniques were originally introduced in the context of High Energy Physics to find localised overdensities in the feature space.
In the case of kinematically cold stellar streams, the member stars are expected to produce localised overdensities in the proper motion feature.
One can define a signal region (SR) based on the proper motion, where there is an increased population of a stellar stream stars, and side bands (SB1, SB2) on either side of the SR, where the stream members are not expected to be present (or at a far lower rate, compared to the SR).

\textbf{\vm} (1.0, 2.0) based on ANODE~\citep{anode}, consists of conditional generative models, that learns the probability distribution of the kinematic, photometric, and astrometric features of the stars in the SR and SB, and constructs the likelihood ratio $\frac{P_{\mathrm{SR}}(\mathbf{x})}{P_{\mathrm{bg}}(\mathbf{x})}$ to tag \textit{anomalous} stars.
Here, $P_{\mathrm{SR}}$ is the probability distribution of the stars in the signal region and $P_{\mathrm{bg}}$ is the conditionally interpolated background density, and $\mathbf{x}$ is the feature vector of the stars, over which the densities are defined.
Thereafter, a line finding algorithm is used to filter out the stream.
The \vm method has been shown to be very effective at finding streams, however, is computationally expensive, as it requires training two generative models on the data.

\textbf{\cwola}, originally introduced in~\citet{cwola}, is a computationally lightweight method that uses a weakly supervised learning approach to detect streams.
Given the SR and SB, \cwola trains a classifier to distinguish between the two regions, and then uses the classifier to tag the stars in the SR.
The classifier effectively learns the likelihood ratio $\frac{P_{\mathrm{SR}}(\mathbf{x})}{P_{\mathrm{SB}}(\mathbf{x})}$.
Here, $P_{\mathrm{SR}}$ is the probability distribution of the stars in the signal region and $P_{\mathrm{SB}}$ is the probability distribution of the stars in the side bands.
However, the performance of \cwola is dependent on the choice of features used in the training.
If the selected features are correlated with the proper motion, the classifier may be biased and produce false excesses in the SR, even in the absence of a stream.

It is possible to circumvent this bias if a suitable template of the background is constructed to be used in the \cwola method.
We propose \textbf{\skyc}, that constructs a background-enriched template of the stars in the SR in a data driven manner.
\skyc is based on \FfF, a method originally developed for anomaly detection in High Energy Physics introduced in ~\citep{curtains,curtainsf4f}.
\FfF is a data-driven weakly supervised strategy that extends the \cwola method to mitigate the problem of correlation of discriminatory features with the proper motion feature.
We leverage the correlation of the features with the proper motion to generate a template in the signal region using the sidebands.
This alleviates the need to sample data from the SB for \cwola, and results in a template that is more representative of the background in the SR.
One can then use the \cwola method to tag the stars in the SR by training a classifier on the template of the SR data, followed by a line finding algorithm to identify the stream.

Constructing a background enriched template significantly reduces false positives, which is a big advantage of the \skyc method over the standalone \cwola method.
As we will see in~\autoref{sec:method} \skyc has a modular design, and its data efficiency in training allows for an efficient scaling of the method to larger number of patches.

\section{Dataset}
\label{sec:gaia_dr2_data}
We demonstrate the \skyc method on the \gaia Data Release 2 (GDR2)~\citep{gdr2} dataset.
GDR2 contains detailed astrometric and photometric information for over 1.3 billion sources in the Milky Way galaxy.
The dataset characterises the source by the right ascension (\ra) and declination (\dec), the parallax (\parallax), proper motions in right ascension (\pmra) and declination (\pmdec), the apparent magnitude (\magnitude), and the colour information in the form of the GBP and GRP bands (\gbprp).
The newer \gaia Data Release 3 (GDR3) comes with improved measurements on radial velocities, but as the \skyc method does not utilise this information, we use the GDR2 dataset.
This allows for a direct comparison with the \vm and \cwola methods, which were developed using the GDR2 dataset.


We use the \gdone as the main stream candidate to benchmark and validate the \skyc method.
The \gdone is a long and dense stream in the Milky Way galaxy, discovered in 2006 by~\citet{gdonediscovery}.
The \skyc method uses the stellar membership of the \gdone in ~\citep{gdone_label_paper}, hereafter referred to as PWB18, as the ground truth to validate the method.
These studies use selections in position, proper motion, colour, and magnitude space to identify the stars that are members of the \gdone.
Although these membership labels are likely not complete and can not be considered as ground truth, they nonetheless provide a crucial reference for the validation of the \skyc method.

Following~\citep{vm1,vm2,astrocwola}, we choose to divide the GDR2 dataset into overlapping circular \textit{patches} of $15^{\circ}$ radius.
We re-center each patch and use \textit{patch local coordinates} (\lon, \lat) and the corresponding proper-motion (\pmlon, \pmlat), such that each patch is centred at $(\alpha_0, \delta_0)$ = $(0^{\circ},0^{\circ})$.
This is done to ensure that each patch has an  approximately Euclidean distance metric.
\begin{figure*}
    \centering
    \includegraphics[width=1.\textwidth]{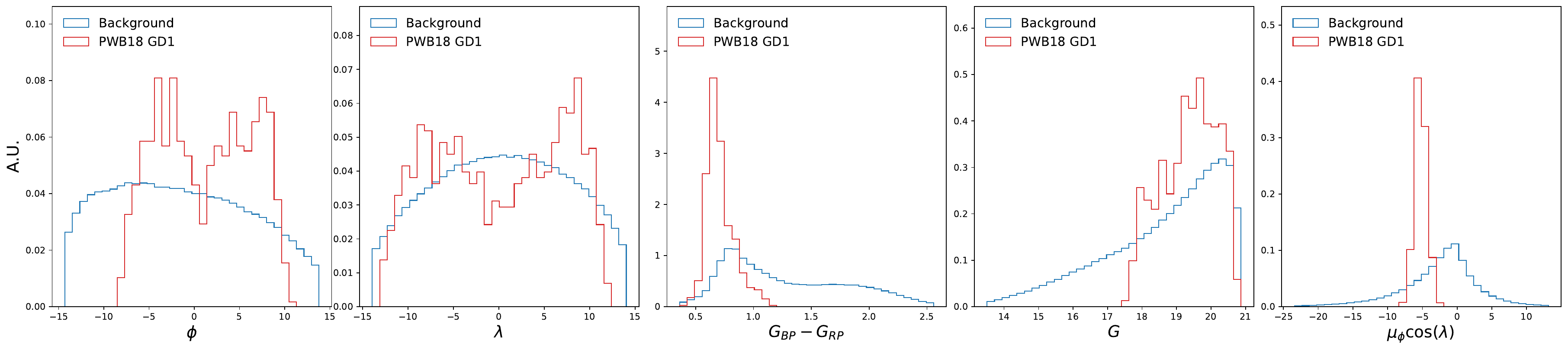}
    \caption{Marginal distribution of the features used in the \skyc method.
    The stars identified by PWB18 study, are shown in red, and the background stars are shown in blue.
    All distributions are normalised to 1, to help visualise the qualitative differences between the features of background and stream like stars.
    The stars in the plot are from the patch with coordinates centred at \ra = $146.9^\circ$, and \dec = $35.6^\circ$}
    \label{fig:feature_space}
\end{figure*}

\skyc uses six features associated with each star: \textit{kinematic} (\pmlonstar = \pmloncoslat, \pmlat), \textit{spatial} (\lon, \lat), and \textit{photometric} (\magnitude, \gbprp).
The marginal distribution of these features is shown in~\autoref{fig:feature_space}.
The entire patch is used to train the \FfF model.
However, additional fiducial cuts are applied to the data for the downstream tasks.
These include:
\begin{itemize}
    \item Kinematic cut: $|\pmlonstar| > 2$ mas/yr OR $|\pmlat| > 2$ mas/yr.
    \item Photometric cut: $0.5 \leq \gbprp \leq 1$, and $\magnitude < 20.2$.
\end{itemize}
The kinematic cuts are applied to reject distant stars that produce an overdensity in the proper motion at $\sim 0$ mas/yr and reduce the sensitivity of the model to overdensities produced by stellar streams.
The cut on magnitude removes stars that are too dim and ensures we have a uniform coverage of stars from the \gaia dataset.
The cut on colour isolates older, low-metallicity stars, which are more likely to be stellar stream members.

\section{\skyc Method}
\label{sec:method}

\skyc is a two stage approach to find stellar streams in a model agnostic manner.
The first stage is \FfF followed by a \cwola step.
This stage is used to infer a threshold to select the candidate signal stars for the second stage.
The first stage flags all overdensities as anomalous.
But as we are looking for stellar streams, we need to identify line-like structures in the candidate signal stars' population.
This is done by the second stage of the \skyc method, which uses the Hough transform~\citep{hough1962} for line detection.
Details of training and implementation of the two stages are discussed in the following sections.

\subsection{\FfF}
\label{sec:curtains}
\FfF constructs a background-enriched template in the signal region by learning a conditional transformation of the features from the sidebands to the signal region, as a function of the proper motion.
\FfF uses a maximum likelihood loss on the transported data and the target data using the \flowsforflows method introduced in~\cite{flows4flows2} to learn this transformation.

A normalising flow~\citep{nflows} is a model that learns a bijective transformation $f_\phi: z \rightarrow x$ between a base distribution to a target distribution under maximum likelihood, where $z \sim p_{\theta}$ and $x \sim P_X$.
The usual choice for the base distribution is a standard normal distribution.
The loss function for training this normalizing flow $f_\phi$ is given by the change of variables formula
\begin{equation*}
    \log P_{\theta, \phi} (x) = \log p_\theta (f_\phi^{-1}(x)) - \log \left| \det (J_{f_\phi^{-1}(x)}) \right|,
\end{equation*}
where $J$ is the Jacobian of $f_\phi$.
In the conditional case this extends to
\begin{equation}
    \label{eq:mle_loss}
    \log P_{\theta, \phi} (x | c) = \log p_\theta \left(f_{\phi}^{-1}(x | c)\right) - \log \left| \det (J_{f_{\phi}^{-1}(x | c)}) \right|,
\end{equation}
where $c$ are the conditional properties, and $\phi$ are the learnable parameters of the normalizing flow $f$, and $\theta$ are the parameters of the base distribution.

\begin{figure}
    \centering
    \includegraphics[width=\columnwidth]{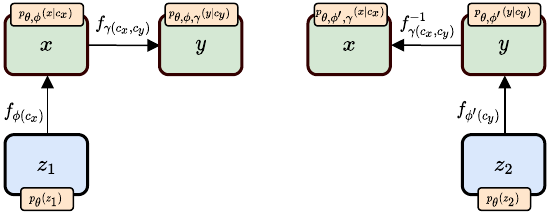}
    \caption{The Flows for Flows architecture for a conditional model.
    Data $x$ ($y$) are drawn from the initial distribution with conditional values $c_x$ ($c_y$) and transformed to new values $c_y$ ($c_x)$ in a cINN $f_{\gamma(c_x,c_y)}$ conditioned on $c_x$ and $c_y$.
    The probability of the transformed data points are evaluated using a second normalizing flow for the base distribution $f_{\phi^\prime(c_y)}$ ($f_{\phi(c_x)}$).
    $Z_1, \mathrm{and} \, Z_2$ denote analytically known distributions, for example, a standard normal.
    In the case where $x$ and $y$ are drawn from the same underlying distribution $p(x, c)$, the same base distribution $f_\phi$ can be used.
    }
    \label{fig:curtainsfff}
\end{figure}

In anomaly detection methods such as \citep{anode}, one learns the distribution  $p_\phi(x|c)$ for $c$ in the sideband regions, and then queries the conditional normalizing flow for $c$ in the signal region to obtain a data-driven model for the background template there.
This ``automatic" interpolation of the conditional density is simple and effective, however empirically it was found in \citep{anode} that for accurate interpolation into the sideband one needed to train $p_\phi(x|c)$ on the entire complement of the signal region. 
For a sliding window search, it is computationally expensive to train a separate flow on the complement of every signal region. \FfF\ improves on this situation by
training a {\it second} conditional flow to learn a transformation between left and right sideband data. 
This flow is found to interpolate much better as the transformations to be learnt are much simpler and this simplicity acts as an implicit regularisation when interpolating to the signal region.
One can get an accurate background template in the signal region with just training on {\it narrower} sidebands instead of the entire complement of the signal region.
The procedure of \FfF\ also allows one to train a single {\it base flow} to learn $p_\phi(x|c)$ for the entire data, and then sampling from this in narrow sidebands one can train the {\it top flow} to interpolate into any signal region.
Thus, the expensive step of training the base flow need only be done once, and then the cheap step of training the top flow can be repeated with much less computational cost.

\begin{subequations}
    \label{eq:fff}
    \begin{align}
        &\max_\gamma \mathop{\mathbb{E}}_{x_1, x_2 \sim P_{\mathrm{SB}}} \left[ \log P_{\theta, \phi, \gamma}(x_1) \right] = \notag\\
        &\max_\gamma \left( \mathop{\mathbb{E}}_{x_1, x_2\sim P_{\mathrm{SB}}} \left[  \log P_{\theta, \phi}(x_2) - \log  \left| \det (J_{f_{\gamma}^{-1}(x_1 | c_{x_1}, c_{x_2})}) \right| \right] \right) \label{eq:topflow};\\
        &\max_\phi \mathop{\mathbb{E}}_{x \sim P_{SB}} \left[ \log P_{\theta, \phi}(x) \right] = \notag\\
        &\max_\phi \left( \mathop{\mathbb{E}}_{x \sim P_{\mathrm{SB}}} \left[\log p_{\theta}(f_{\phi}^{-1}(x | c_x)) - \log  \left| \det (J_{f_{\phi}^{-1}(x | c_x)}) \right| \right] \right) \label{eq:baseflow}.
    \end{align}
\end{subequations}
The optimisation problem for \FfF is shown in~\autoref{eq:fff}.
~\autoref{eq:topflow} pertains to the optimisation of the top flow $f_{\gamma}$, while~\autoref{eq:baseflow} pertains to the optimisation of the base flow $f_{\phi}$ that defines the log-likelihood term in the first equation.
where $p_{\mathrm{SB}}$ denotes the distribution of features in the sidebands.
$f_\gamma^{-1}(x_1 | c_{x_1}, c_{x_2}) = x_2$ is the conditional top flow, where $x_1$ and $x_2$ are drawn from the sidebands, and $c_{x_1}$ and $c_{x_2}$ are the conditional properties of $x_1$ and $x_2$ respectively.
$f_\phi^{-1}(x | c_x) = z$ is the conditional base flow, where $z \sim p_{\theta}(z)$ is drawn from a standard normal distribution.
The correspondence between the top normalizing flow and the base distributions in Flows for Flows is shown in~\autoref{fig:curtainsfff}.

\begin{figure}
    \centering
    \includegraphics[width=0.9\columnwidth]{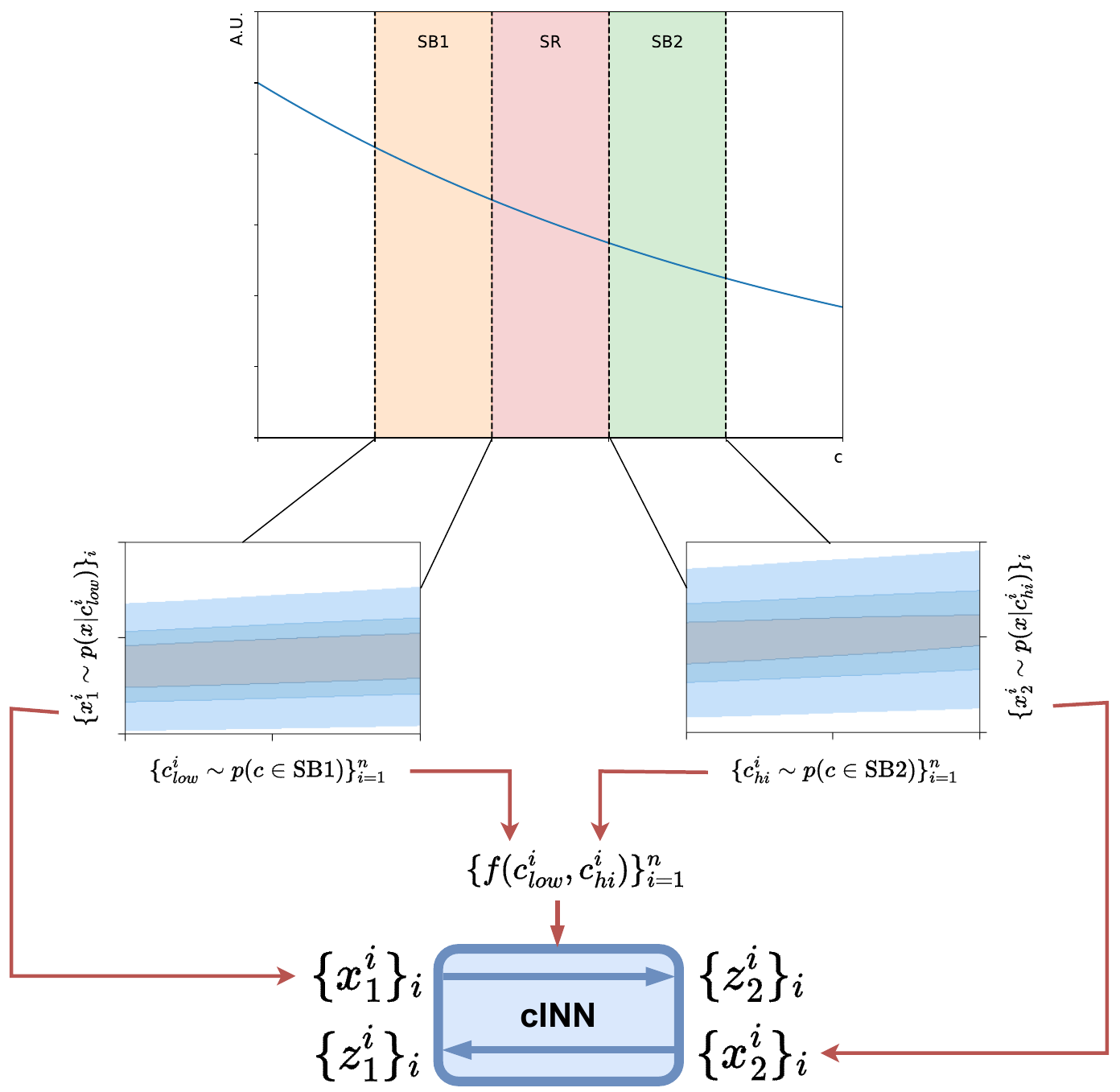}
    \caption{A schematic overview of the training procedure for \FfF with an event where the target proper motion (c) value is greater than the input value.
    A single conditional normalizing flow is used for the base distribution, conditioned on the target c value $c_\textit{target}$, to determine $p_\theta(z|c_\textit{target})$.
    The top normalizing flow is conditioned on a function of the input ($c_\textit{input}$) and target ($c_\textit{target}$) proper motion values.
    For the case where $c_\textit{target} < c_\textit{input}$, an inverse pass of the network is used, and the conditioning property is calculated as $f(c_\textit{target} , c_\textit{input})$.}
    \label{fig:schema}
\end{figure}

\FfF is trained in both directions.
The forward pass transforms data from low to higher target values of proper motion, whereas the inverse pass transforms data from high to lower target values.
Data are drawn from both SBs and target proper motion values are randomly assigned to each data point using all proper motion values in the batch.
Data are passed through the network in a forward or inverse pass, depending on whether the proper motion is larger or smaller than their initial proper motion.
The network is conditioned on a function of initial and target proper motion, with the two values ordered in ascending order.
This function could be, for example, difference between the two, or simply both values concatenated.
The probability term is evaluated using a single base distribution trained on the data from \SBone and \SBtwo.
The loss for the batch is calculated from the average of the probabilities calculated from the forward and inverse passes.
A schematic overview is shown in~\autoref{fig:schema}.

The base flow is trained on the sideband data with a standard normal distribution as the target prior.
It is conditioned on the proper motion.
The top flow is trained between data drawn from the sidebands.
The transformation is conditioned on the concatenated tuple of the initial and target proper motion.
Depending on which proper motion is chosen as the conditioning feature, the downstream task of finding the stream might be affected.
This is because the stream candidate stars may have a non-trivial correlation, and therefore may produce overdensities of different shapes in the two proper motions respectively.
In this work, we use \pmlat as the conditional feature.
The rest of features, i.e. [\lon, \lat, \magnitude, \gbprp, \pmloncoslat] are used to characterise the template.

\begin{figure}
    \centering
    \includegraphics[width=0.9\columnwidth]{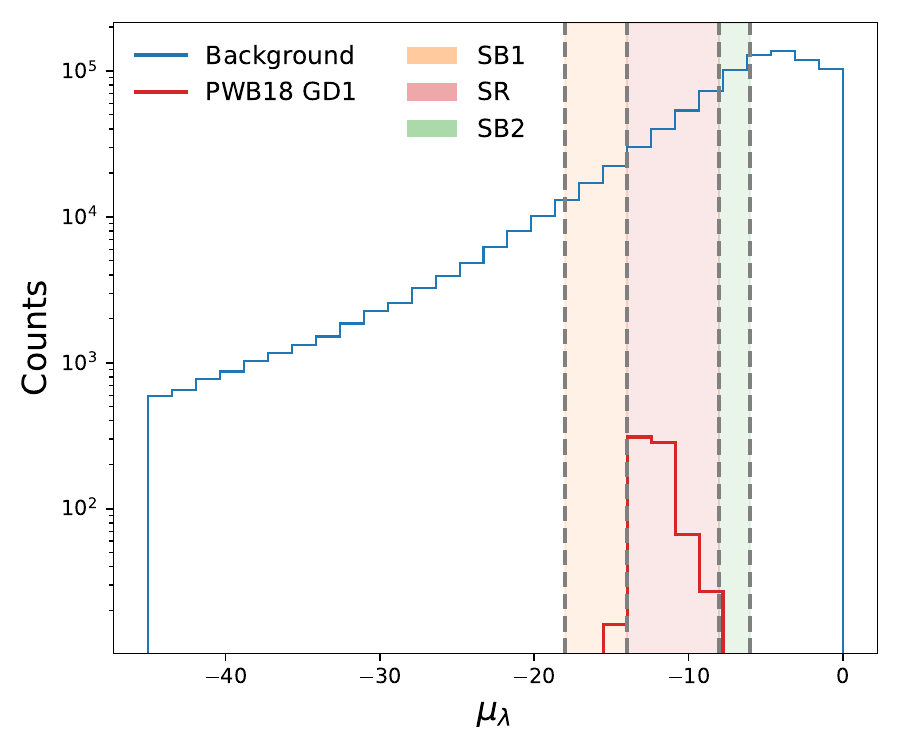}
    \caption{Distribution of \pmlat of background stars (in blue) and \gdone stars (in red) in the sidebands and signal region defined in bins of \pmlat.
    The signal region is defined as \pmlat $\in \left[-14, -8\right)$, and the sidebands are defined as \pmlat $\in \left[-18, -14\right)$ and \pmlat $\in \left[-8, -6\right)$.
    In this SR, there are 184142 background like stars, and 688 PWB18 tagged \gdone stars, which is a signal fraction of only $0.0037$.
    }
    \label{fig:windows}
\end{figure}

One important aspect of the \FfF method is the definition of signal and sideband regions.
In~\autoref{fig:windows}, we show the distribution of \pmlat of the background stars and \gdone stars in the sidebands (\SBone, \SBtwo) and signal region (SR).
Unlike in the \vm method, where the sideband region was the complementary region of the chosen SR, \skyc defines the sideband region to be typically of 2-6 mas/yr.
Since the top flow only needs to learn a small (but not necessarily trivial) transformation of the sideband data to generate a template in the SR, we find this width of the sideband to be sufficient.
This also cuts down the total training time compared to \vm, as despite both methods consisting of two generative models, \skyc's second generative model effectively learns on narrower sidebands.
To demonstrate the efficacy of the method on the \gdone, we define the signal region as the interval in which the signal is contained.
In an actual analysis, where the location of the signal is not known a priori, one would need to scan multiple values of \pmlat with the \FfF method.
Here, the modularity of the \FfF method comes into play, as the base flow can be trained on the entire patch of the sky and frozen.
Thereafter, top flows can be trained on individual regions of interest.
This significantly reduces the computational cost of training the model, by allowing the base flow to be trained once and reused for multiple regions of interest.

Once the \FfF model is trained, the background-enriched template is constructed by transforming the data from the sidebands to the signal region, conditioned on the tuple of the initial and target proper motion.
The target proper motions in the signal region are sampled from a kernel density fit on the proper motion distribution.
We train an ensemble of $10$ multi layer perceptron based classifiers\footnote{Details about the architecture can be found in Appendix~\ref{sec:appendix_tuning}} between this template and the signal region data.
The scores are then aggregated by taking the mean score and used to classify the data in the signal region.
The top 0.1\% most signal-like stars are selected as the candidates for the next step.
This threshold is selected to remove the most background like stars from the signal region, which is a tunable parameter and can be adjusted based on the desired purity and signal efficiency.

\subsection{Line detection}
\label{sec:hough}
The \FfF step gives us a set of stars which produce an overdensity in the feature space.
We still need to filter out the overdensities that are particularly line like, as we are interested in stellar streams.
We employ a well known line finding algorithm to estimate the line parameters of the stream via the Hough transform, as was done in~\citep{vm1,vm2}.
Since we do not apply a fiducial cut to eliminate stars outside a $10^{\circ}$ radius, we can use the full set of stars that pass the \FfF cut in the patch to estimate the line parameters.

For a given star located at (\lon$'$, \lat$'$), the Hough transform is a mapping from the \lon-\lat space to the parameter space $\rho-\theta$, defined as:
\begin{equation}
    \label{eq:hough_transform}
    \rho = \lon' \cos \theta - \lat' \sin \theta
\end{equation}
where $\rho$ is the perpendicular distance from the origin to the line and $\theta$ is the angle between the line passing through (\lon$'$, \lat$'$) and the \lon-axis.
The origin is defined as the point $(0,0)$ in the \lon-\lat space.
A line in the \lon-\lat space is represented as a point in the $\rho-\theta$ space.
If a set of points lie on a line in the \lon-\lat space, they will map to a single point in the $\rho-\theta$ space.
Therefore, identifying line candidates in the \lon-\lat space is equivalent to identifying high density regions in the $\rho-\theta$ space.
For each patch, we bin the Hough space in a $100 \times 100$ grid from $-15^{\circ} \leq \rho \leq 15^{\circ}$ and $0 \leq \theta \leq \pi$, such that each bin $(i,j) \left( i \in [0, 99], j \in [0, 99] \right)$ is related to the $\rho$ and $\theta$ values by:
\begin{eqnarray}
    \rho_i &=& -15 +  i  \times \frac{30}{99} \\
    \theta_j &=& \pi \times \frac{j}{99}
\end{eqnarray}

\begin{figure*}
    \centering
    \begin{subfigure}{\columnwidth}
        \centering
        \includegraphics[width=\columnwidth]{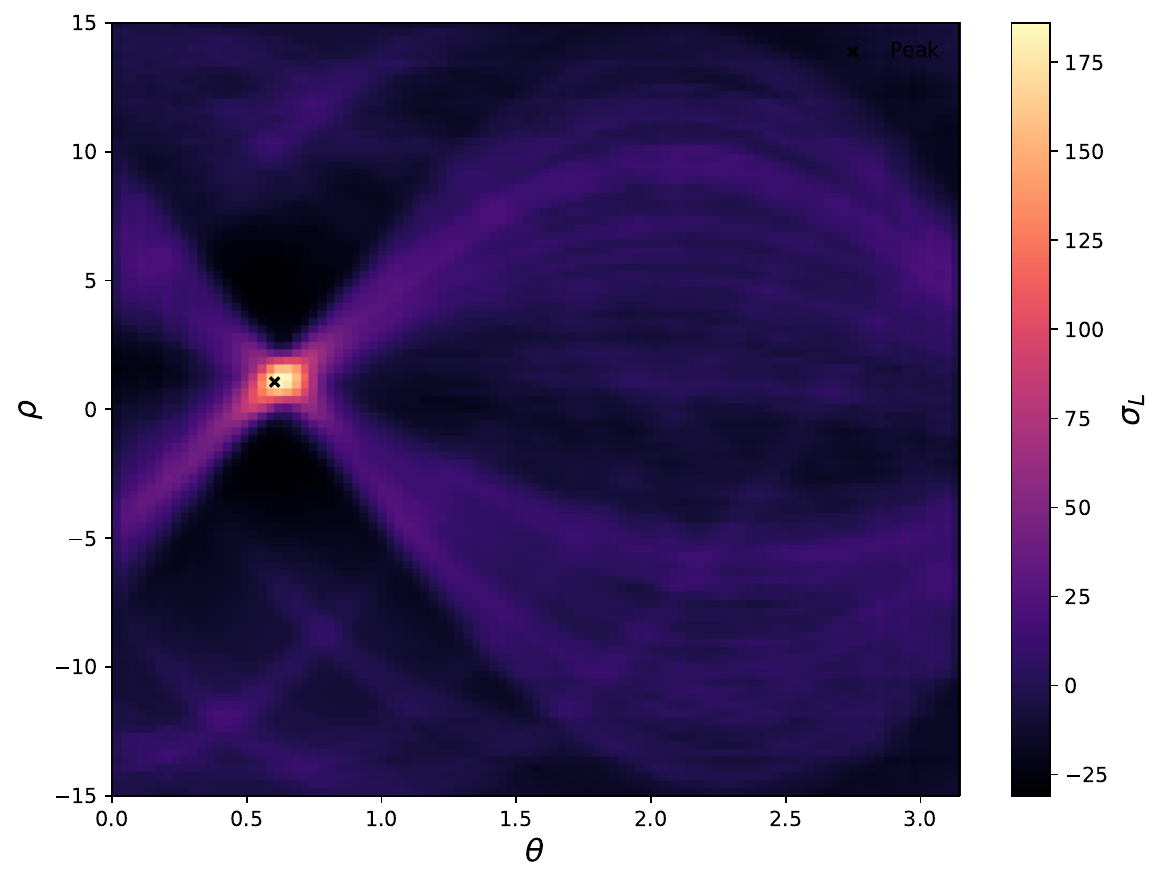}
    \end{subfigure} %
    \begin{subfigure}{\columnwidth}
        \centering
        \includegraphics[width=\columnwidth]{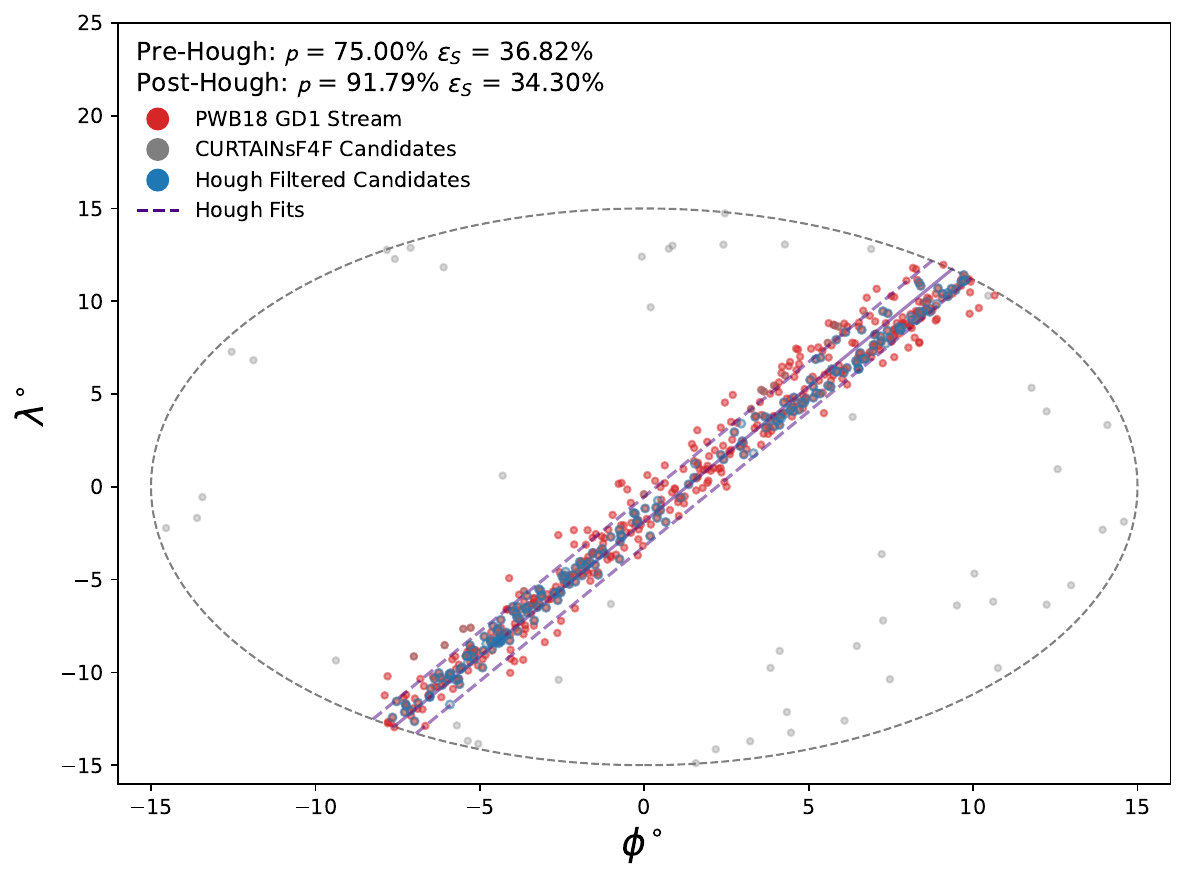}
    \end{subfigure}
    \caption{Left: The significance map of the Hough space for candidate signal stars in the patch with coordinates \ra = $146.9^\circ$, and \dec = $35.6^\circ$. 
    Each pixel represents the number of stars whose Hough curve passes through a box of width $(\Delta \rho = 1.5^{\circ}, \Delta \theta = 0.03 \, \mathrm{rad})$ centred at that pixel.
    The bright spots correspond to the peaks in the Hough space.
    The highest significance pixel is marked with a black cross.
    Right: \lon-\lat scatter plot of stars post \cwola step in the same patch.
    The red coloured stars correspond to the \gdone as identified by PWB18.
    The grey coloured stars are those selected by \skyc as the most signal like stars in this patch.
    The blue coloured stars are those selected after applying a hough filter and fall within the acceptance region defined by the Hough fit lines in purple.}%
    \label{fig:hough_stream}
\end{figure*}

Stellar streams have a finite width in space.
To account for this, we calculate the number of lines passing through a box of width ($\Delta i =5, \Delta j = 3$) centred at $(i, j)$.
This is done by convolving the Hough space with a uniform kernel of size $(\Delta i, \Delta j)$.
These widths correspond to $\Delta \rho = 1.5^{\circ}$ and $\Delta \theta = 0.09$ rad.
Thereafter, similar to~\citep{vm2}, we scan this convolved Hough space for high significance peaks.
This allows us to calculate a range of $\rho$ and $\theta$ values that correspond to the line-like structures in the \lon-\lat space, and account for the finite width of the stream.
~\autoref{fig:hough_stream} (left) shows the Hough space for the \gdone in one of the patches.

\section{Results}
The \FfF stage was trained on \nvidia RTX 3080 GPUs, and the Hough stage was run on a single CPU core.
The \FfF stage took $\sim 4$ hours per patch, amounting to a total of $\sim 80$ GPU hours.
The \FfF stage consists of training a base and a top flow.
Both base and top flow took $\sim 2$ hours to train.
The line fitting took about a minute per patch, and was a negligible fraction of the total computational cost.

\label{sec:results}
\begin{figure*}
    \centering
    \begin{subfigure}{\columnwidth}
        \centering
        \includegraphics[width=\columnwidth]{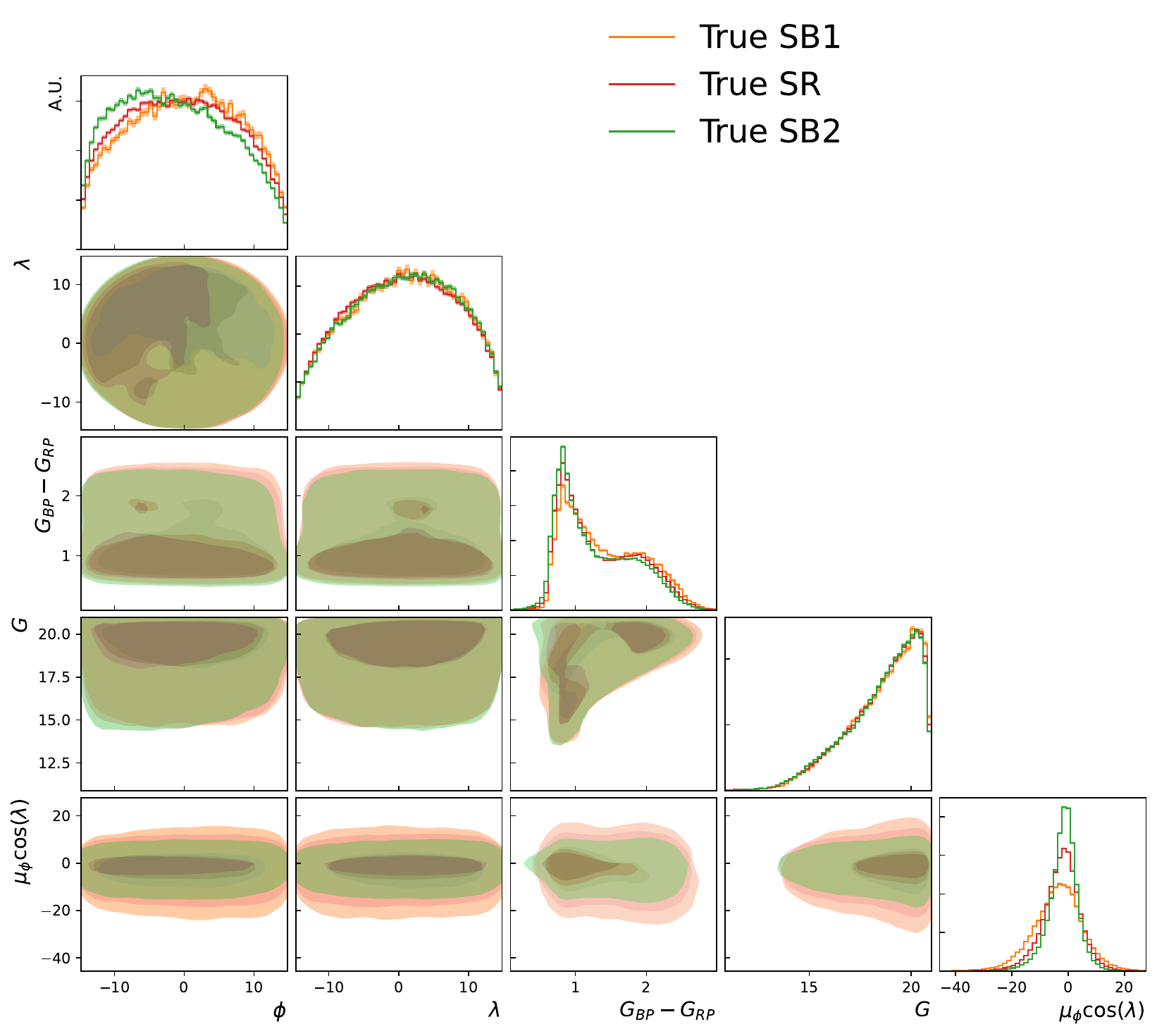}
    \end{subfigure}
    \begin{subfigure}{\columnwidth}
        \centering
        \includegraphics[width=\columnwidth]{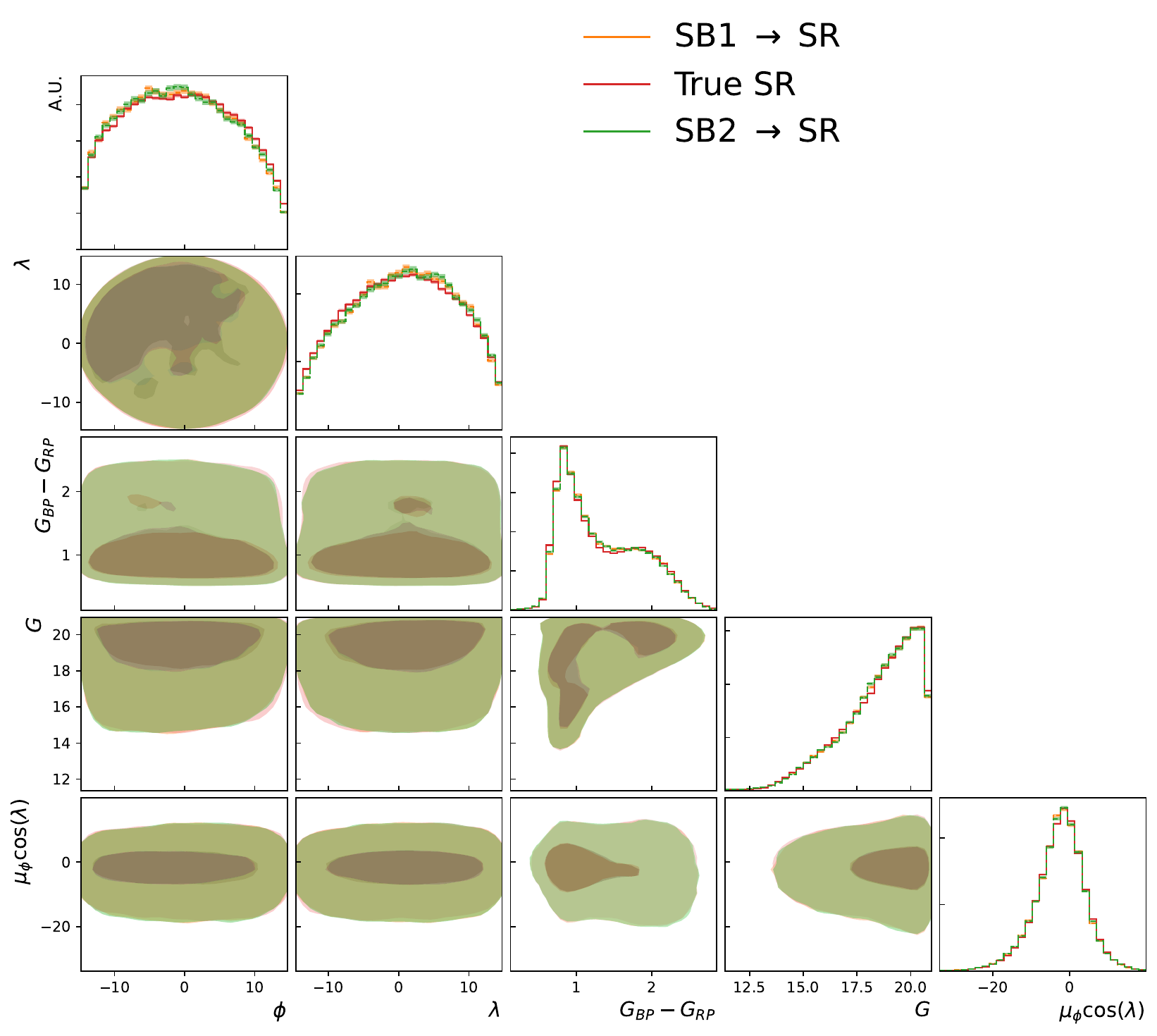}
    \end{subfigure}
    \caption{Feature correlation plots in SB1 (orange), SR (red), and SB2 (green).
    The diagonal panels show the marginals of the features in the SB1 (orange), SR (red), and SB2 (green).
    The off-diagonal panels show the correlation between the features.
    The left panel shows the feature correlation plots of true sideband region data and true signal region data.
    The right panel shows the feature correlation plots of the \FfF generated template (\SBone $\rightarrow$ SR, \SBtwo $\rightarrow$ SR), and true signal region data.
    The generated template has a much better agreement with the true signal region data, and preserves the correlation between the features.
    The features correspond to the patch centred at \ra = $146.9^\circ$, and \dec = $35.6^\circ$}.%
    \label{fig:template}
\end{figure*}

The most crucial step in the \skyc method is the generation of a background enriched template in the signal region.
In~\autoref{fig:template}, we show the marginals and correlations of features in the sidebands and signal region in the left panel.
The features are strongly correlated with the proper motion, which would bias the classifier in the \cwola step to produce false positives in the signal region even in the absence of a stream.
In the right panel, we show the marginals and correlations of the features in the generated template by \FfF in the same patch.
The generated template leverages the correlation of the features with the proper motion to construct a background enriched template in the signal region.
This allows for a more representative template of the background in the signal region, and reduces the false positives in the search for stellar streams.
With the generated template, we can now train a classifier in the \cwola step to tag the stars in the signal region.

\subsection{Metrics}
\begin{figure}
    \centering
    \includegraphics[width=\columnwidth]{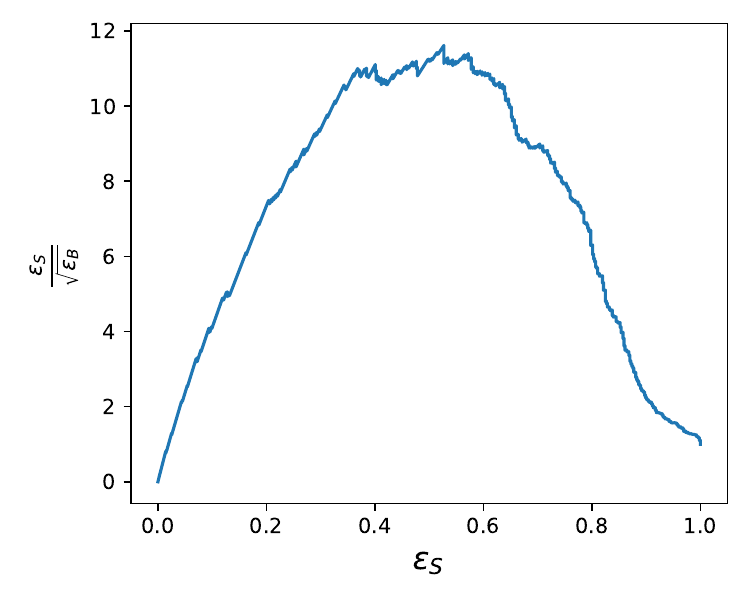}
    \caption{Significance improvement characteristic curve as a function of signal efficiency for the \gdone  in the patch with coordinates \ra = $146.9^\circ$, and \dec = $35.6^\circ$.}%
    \label{fig:sicsvb}
\end{figure}

We now demonstrate the performance of the \skyc method on GDR2 data.
To quantify the discovery potential of \skyc method, we measure the Significance Improvement Characteristic (SIC) curve for the \gdone.
In~\autoref{fig:sicsvb}, we show the SIC curve as a function of the signal efficiency for the \gdone in one of the 21 patches.
This metric is defined as the ratio of the signal efficiency to the square root of the background efficiency, and essentially quantifies the improvement in the discovery significance of the signal from the method.
\skyc achieves a maximum significance improvement of $\sim 10$ at $\sim 50\%$ signal efficiency.
Although a direct comparison with \vm is difficult on account of different SR being used for the analysis, one can look at the maximum value of the SIC as a heuristic measure, which are comparable for both methods.

We track two other metrics to quantify the performance of \skyc: \textit{purity} $p$: The fraction of candidate \FfF stars that overlap with the PWB18 identified \gdone members; and signal efficiency, $\epsilon_S$ which is the fraction of \gdone members that have been flagged as candidates by \FfF step.
~\autoref{fig:hough_stream} (right) shows the candidates from the \FfF step in the \lon-\lat space that corresponds to the \gdone with a $p = 75 \%$ and $\epsilon_S = 36.82 \%$.
We note that it also predicts a few stars that do not form a line like structure in the \lon-\lat space.
This is expected, as this stage is designed to flag any overdensity in the feature space as a potential signal candidate.
To filter out the line like overdensities we perform a Hough transform on the output of \FfF step.
After applying the Hough filter, the purity is improved to $91.79\%$, albeit at the cost of a slightly reduced signal efficiency of $34.3\%$.
\subsection{Full \gdone scan}
\label{sec:patchperf}

\begin{figure*}
    \centering
    \begin{subfigure}{1.0\textwidth}%
        \centering
        \includegraphics[width=\textwidth]{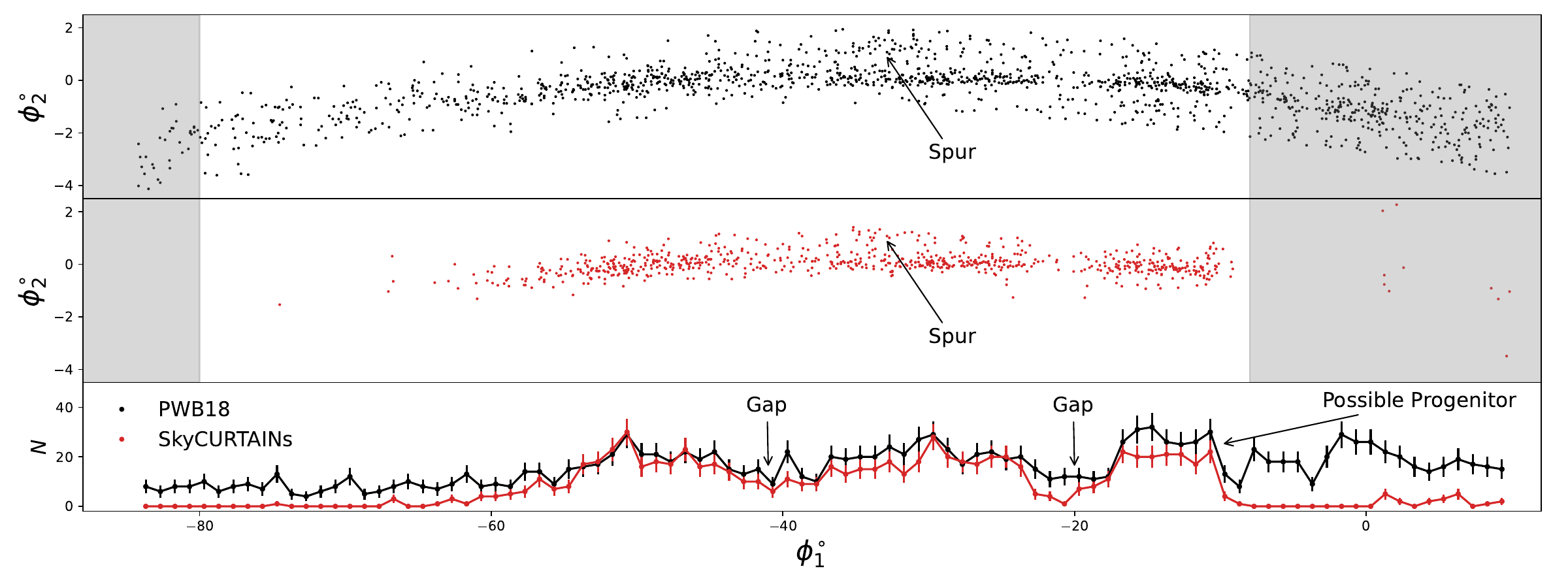}
    \end{subfigure} \\
    \begin{subfigure}{0.5\textwidth}

        \includegraphics[width=\columnwidth]{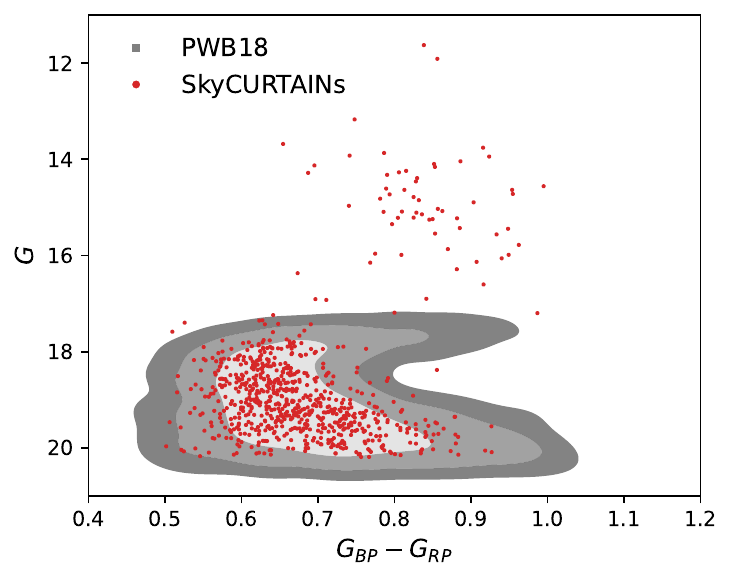}
    \end{subfigure}%
    \begin{subfigure}{0.5\textwidth}
        \includegraphics[width=\columnwidth]{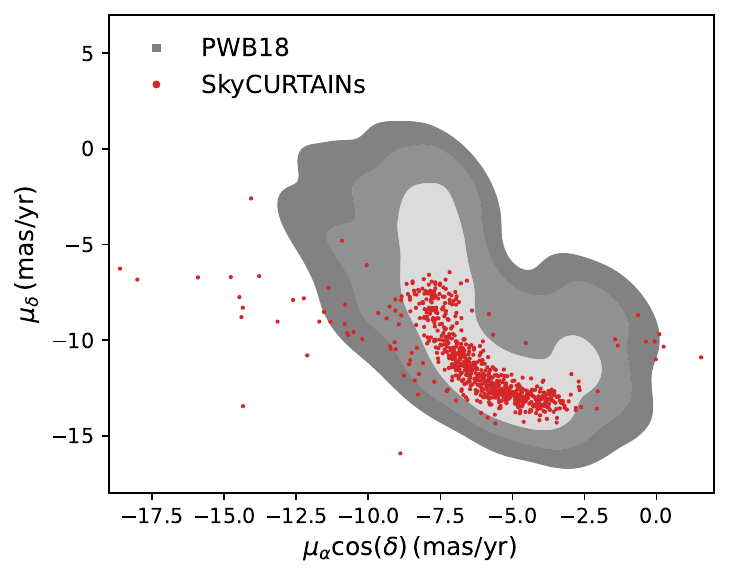}
    \end{subfigure}
    \caption{Top: Comparison of PWB18 and the 753 \skyc identified stream candidates.
    The first panel shows the PWB18 \gdone members in the \gdone stream aligned coordinate system $(\phi_1, \phi_2)$.
    The middle panel shows the potential members of the \gdone stream in the \gdone aligned coordinate system identified by the \skyc method.
    The bottom panel is a comparison of the number of candidate stream stars identified by the \skyc method (red) and the number of PWB18 (black) in the \gdone in $\phi_1$ bins of width $1^{\circ}$.
    The shaded region to the left of the stream correspond to the patches which are excluded from the analysis due to their proximity to the galactic disk.
    The shaded region to the right of the stream correspond to the patches containing stars with very low proper motions, which reduces the signal sensitivity of the \FfF to potential streams.
    Bottom: Coverage plots in the \magnitude vs \gbprp and \pmra vs \pmracosdec space.
    The contours show the 68.2, 95.4 and 99.7 percentiles of the \gdone members identified by PWB18.
    The \skyc identified candidates are shown in red.}
    \label{fig:cmd}
\end{figure*}

~\autoref{tab:patchperf} shows the performance of the \skyc method in the 21 patches that contain the \gdone.
We quote the purity $p$ after applying the Hough filter for each patch.
\skyc is able to identify the \gdone members with a high purity in most of the patches, and significantly improves the performance compared to standalone \cwola.
In~\autoref{tab:patchperfraw} we report the total PWB18 identified \gdone members and \skyc candidates (after Hough filter) in the patches.
The combined result is shown in~\autoref{fig:cmd}.

\begin{table}
    \centering
    \caption{Performance of the \skyc method in the 21 patches that contain the \gdone.
    The patches are identified by the central \ra and \dec of the patch.
    We quote the purity $p$ after applying the Hough filter for each patch, and compare the performance with standalone \cwola.
}
    \label{tab:patchperf}
    \begin{tabular}{lrr}
    \hline 
    \multicolumn{1}{c}{\multirow{2}{*}{\textbf{Patch (\ra, \dec)}}} & \multicolumn{2}{c}{$p$}\\
    \multicolumn{1}{c}{} & \multicolumn{1}{c}{\textbf{\skyc}} & \textbf{\cwola} \\
    \hline 
    \hline 
    $\left(128.4^{\circ}, \,\, 28.8^{\circ}\right)$ &     $\mathbf{82.99}$ &          $77.0$ \\
    $\left(132.6^{\circ}, \,\, 16.9^{\circ}\right)$ &     $\mathbf{78.05}$ &          $62.0$ \\
    $\left(136.5^{\circ}, \,\, 36.1^{\circ}\right)$ &     $\mathbf{90.56}$ &          $86.0$ \\
    $\left(138.8^{\circ}, \,\, 25.1^{\circ}\right)$ &     $\mathbf{90.79}$ &          $84.0$ \\
    $\left(142.7^{\circ}, \,\, 14.5^{\circ}\right)$ &     $\mathbf{86.79}$ &          $65.0$ \\
    $\left(146.9^{\circ}, \,\, 35.6^{\circ}\right)$ &     $\mathbf{91.79}$ &          $90.0$ \\
    $\left(148.6^{\circ}, \,\, 24.2^{\circ}\right)$ &      $\mathbf{94.9}$ &          $87.0$ \\
    $\left(148.6^{\circ}, \,\, 47.0^{\circ}\right)$ &     $\mathbf{93.15}$ &          $78.0$ \\
    $\left(156.2^{\circ}, \,\, 57.5^{\circ}\right)$ &     $\mathbf{70.14}$ &          $54.0$ \\
    $\left(156.9^{\circ}, \,\, 34.1^{\circ}\right)$ &     $\mathbf{88.17}$ &          $86.0$ \\
    $\left(160.5^{\circ}, \,\, 45.5^{\circ}\right)$ &     $\mathbf{87.43}$ &          $73.0$ \\
    $\left(171.4^{\circ}, \,\, 43.0^{\circ}\right)$ &     $\mathbf{89.52}$ &          $72.0$ \\
    $\left(171.8^{\circ}, \,\, 54.7^{\circ}\right)$ &     $\mathbf{89.66}$ &          $53.0$ \\
    $\left(174.3^{\circ}, \,\, 65.1^{\circ}\right)$ &     $\mathbf{64.94}$ &          $47.0$ \\
    $\left(185.4^{\circ}, \,\, 50.0^{\circ}\right)$ &      $\mathbf{84.0}$ &          $57.0$ \\
    $\left(192.0^{\circ}, \,\, 58.7^{\circ}\right)$ &     $\mathbf{83.87}$ &          $66.0$ \\
    $\left(138.1^{\circ}, \,\, 5.7 ^{\circ}\right)$ &                $0.0$ &           $0.0$ \\
    $\left(203.7^{\circ}, \,\, 49.1^{\circ}\right)$ &      $\mathbf{0.13}$ &           $0.0$ \\
    $\left(212.7^{\circ}, \,\, 55.2^{\circ}\right)$ &                $0.0$ &           $0.0$ \\
    $\left(224.7^{\circ}, \,\, 60.6^{\circ}\right)$ &      $\mathbf{2.58}$ &           $0.0$ \\
    $\left(202.4^{\circ}, \,\, 66.5^{\circ}\right)$ &                $0.0$ & $\mathbf{50.0}$ \\
    \hline 
    \end{tabular}
\end{table}

\begin{table}
    \centering
    \caption{PWB18 identified \gdone members and \skyc candidates in the 21 patches that contain the \gdone.
}
    \label{tab:patchperfraw}
    \begin{tabular}{lrr}
    \hline 
    \multicolumn{1}{c}{\multirow{1}{*}{\textbf{Patch (\ra, \dec)}}} & \multicolumn{1}{c}{\textbf{PWB18}} & \multicolumn{1}{c}{\textbf{\skyc}}\\
    \hline 
    \hline 
    $\left(128.4^{\circ}, \,\, 28.8^{\circ}\right)$ & $307$ & $147$ \\
    $\left(132.6^{\circ}, \,\, 16.9^{\circ}\right)$ & $321$ & $41$ \\
    $\left(136.5^{\circ}, \,\, 36.1^{\circ}\right)$ & $428$ & $180$ \\
    $\left(138.8^{\circ}, \,\, 25.1^{\circ}\right)$ & $421$ & $76$ \\
    $\left(142.7^{\circ}, \,\, 14.5^{\circ}\right)$ & $312$ & $106$ \\
    $\left(146.9^{\circ}, \,\, 35.6^{\circ}\right)$ & $564$ & $207$ \\
    $\left(148.6^{\circ}, \,\, 47.0^{\circ}\right)$ & $470$ & $73$ \\
    $\left(148.6^{\circ}, \,\, 24.2^{\circ}\right)$ & $415$ & $98$ \\
    $\left(156.2^{\circ}, \,\, 57.5^{\circ}\right)$ & $473$ & $144$ \\
    $\left(156.9^{\circ}, \,\, 34.1^{\circ}\right)$ & $541$ & $186$ \\
    $\left(160.5^{\circ}, \,\, 45.5^{\circ}\right)$ & $584$ & $167$ \\
    $\left(171.4^{\circ}, \,\, 43.0^{\circ}\right)$ & $551$ & $229$ \\
    $\left(171.8^{\circ}, \,\, 54.7^{\circ}\right)$ & $585$ & $116$ \\
    $\left(174.3^{\circ}, \,\, 65.1^{\circ}\right)$ & $453$ & $77$ \\
    $\left(185.4^{\circ}, \,\, 50.0^{\circ}\right)$ & $551$ & $50$ \\
    $\left(192.0^{\circ}, \,\, 58.7^{\circ}\right)$ & $583$ & $62$ \\
    $\left(138.1^{\circ}, \,\,  5.7^{\circ}\right)$ & $209$ & $3$ \\
    $\left(203.7^{\circ}, \,\, 49.1^{\circ}\right)$ & $380$ & $4$ \\
    $\left(212.7^{\circ}, \,\, 55.2^{\circ}\right)$ & $336$ & $0$ \\
    $\left(224.7^{\circ}, \,\, 60.6^{\circ}\right)$ & $244$ & $28$ \\
    $\left(202.4^{\circ}, \,\, 66.5^{\circ}\right)$ & $380$ & $0$ \\
    
    \hline 
    \end{tabular}
    \end{table}

\begin{figure}
    \centering
    \includegraphics[width=\columnwidth]{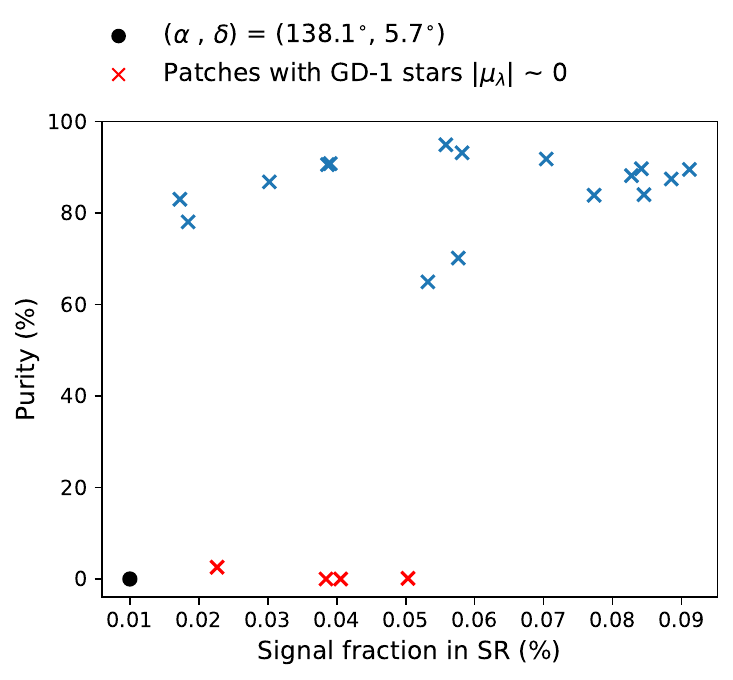}
    \caption{\gdone purity obtained by the \skyc method as a function of PWB18 signal to background ratio (shown here in percentages).
    }
    \label{fig:patch_sr}
\end{figure}

\skyc has a very low purity in 5 of the 21 patches of the sky where the \gdone is present.
On closer inspection we find that in patches (\ra, \dec) = $\left[ \left(203.7^{\circ}, 49.1^{\circ}\right),\left(212.7^{\circ}, 55.2^{\circ}\right), \left(224.7^{\circ}, 60.6^{\circ}\right), \left(202.4^{\circ},66.5^{\circ}\right) \right]$, the \gdone members peak at very low proper motion (\pmlat).
This results in a SR that is dominated by distant stars, and the sensitivity of the \FfF step to actual stream stars is reduced.
These patches correspond to $\phi_1 \geq -10^{\circ}$ in the \gdone aligned coordinates, which explains the low yield of the \skyc method to the right of the stream in~\autoref{fig:cmd}.
\skyc also has a low purity in the patch centred at (\ra, \dec) = $\left(138.1^{\circ}, 5.7 ^{\circ}\right)$.
The low \gdone purity in this patch is likely due to the extremely low signal to background ratio in the corresponding SR.
In~\autoref{fig:patch_sr} we show the \gdone purity as a function of the PWB18 signal to background ratio.
We find the purity has a sharp drop to zero when the signal to background ratio is near 0.01\%.
Patch (\ra, \dec) = $\left(138.1^{\circ}, 5.7^{\circ}\right)$ has a signal to background ratio of 0.01\%, which is the lowest in the 21 patches.
This patch corresponds to $-80^{\circ} \leq \phi_1 \leq -60^{\circ}$ in the \gdone aligned coordinates, and explains the low yield of the \skyc method to the left of the stream in~\autoref{fig:cmd}.
These patches (marked in red) are the patches where the \gdone members peak at very low proper motion (\pmlat), and the sensitivity of the \FfF step to actual stream stars is reduced.
This patch corresponds to $-80^{\circ} \leq \phi_1 \leq -60^{\circ}$ in the \gdone aligned coordinates, and explains the low yield of the \skyc method to the left of the stream in~\autoref{fig:cmd}.

It is crucial to note that \skyc method assumes very little astrophysical information about the stream, allowing it to be agnostic to the stream's properties.
The only information used in the method is the proper motion which is used to define the SR and SB regions.
For stream identification, fiducial cuts on \gbprp and \magnitude (there no requirements on streams to lie on an isochrone) are applied.
This is in parity with the fiducial cuts applied in~\citep{astrocwola, vm1}.
\skyc flags 753 unique stars as potential \gdone members, of which 568 are also identified by PWB18, thereby attaining an overall \gdone purity of $75.4\%$.
This surpasses the standalone \cwola method which has a purity of $56\%$, and \vm 1.0 which has a purity of $49\%$.
\skyc also outperforms \vm 2.0, which has a purity of $65\%$, despite the latter employing additional fiducial cuts and performs an augmented scan over both proper motions.
There are 1498 PWB18 identified \gdone stars in our fiducial region, which gives us a global signal efficiency of $37.9\%$.
Furthermore, an important result of the \skyc method is that it produces no spurious streams in the 21 patches that were scanned.
This can be attributed to the very stringent selection criteria applied in the \FfF stage of the method, designed to reduce false positives.

\begin{figure}
    \centering
    \includegraphics[width=\columnwidth]{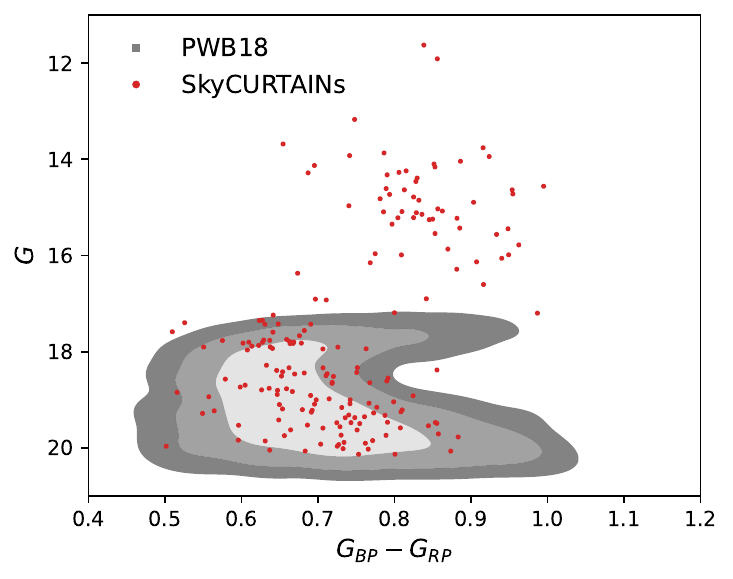}
    \caption{Coverage plot for the \gdone in the \magnitude vs \gbprp space.
    The contours show the 68.2, 95.4 and 99.7 percentiles of the \gdone members identified by PWB18.
    The 185 additional \skyc identified candidates are shown in red.
    }
    \label{fig:colorextra}
\end{figure}

Of the remaining 185 stars, some may potentially be new undiscovered members of the \gdone.
\autoref{fig:colorextra} shows the isochrone plot for the \gdone members identified by PWB18, with the additional \skyc candidates overlaid.
There is a significant overlap between these 185 stars and the PWB18 labelled members, which suggests that the \skyc method is able to identify some members of the \gdone that may have been missed by PWB18.
There are also a few stars that are not part of the \gdone isochrone, and are likely to be false positives.

Despite the lack of prior astrophysical information, the \skyc method is able to recover well known density perturbations in the \gdone.
In the \gdone stream aligned coordinates $(\phi_1, \phi_2)$~\citep{stream_potential3} shown in~\autoref{fig:cmd}, we see that \skyc recovers the "gaps" at $\phi_1 \approx -40^{\circ}$ and $\phi_1 \approx -20^{\circ}$, as well as the "offshoot" or "spur" at $\phi_1 \approx -35^{\circ}$, which are well known features of the \gdone.
Furthermore, \skyc predictions of the overdensity regions at $\phi_1 \approx -50^{\circ}$ and $\phi_1 \approx -10^{\circ}$ are in good agreement with the PWB18 members.
The low yield regions at $\phi_1 \geq -10^{\circ}$ and $-80^{\circ} \leq \phi_1 \leq -60^{\circ}$ are due to the reasons discussed above.
The region $\phi_1 \leq -80^{\circ}$ correspond to the patches that are excluded from the analysis due to their proximity to the galactic disk.
\section{Conclusion}
\label{sec:outlook}

In this work, we described the \skyc method, a model-agnostic, template based, data-driven approach to detecting stellar streams in the Milky Way using the \gaia DR2 data.
Originally developed for anomaly detection in High Energy Physics, \skyc joins the ranks of \vm and \cwola, in the search for stellar streams, which highlights the versatility of these tools in their performance across different domains.
Synergies between the High Energy Physics, Astrophysics, and other communities should be further encouraged in order to identify similar problems that can be solved using the same tools developed in respective fields.

We demonstrated the performance of \skyc on the \gdone, and its ability to identify the line-like overdensity in the \lon-\lat space that corresponds to the \gdone with a very high purity across most patches.
The main advantage of \skyc is the minimal assumptions it makes about the underlying signal, thereby making it a versatile tool for identifying any localized overdensities (streams, globular clusters, dwarf galaxies) in the feature space of the stars in a model agnostic manner.
In this work we chose \gdone, as it provides a good test case for \skyc, and is a well known stream in the Milky Way, where we show that \skyc currently outperforms the other weakly supervised machine learning methods like \vm and \cwola in terms of purity by over 10\%.
For a full sky scan for streams, where the locations of the streams are unknown, the method will need to scan a larger number of \pmlat.
In this case, training two conditional generative models for each patch may not be feasible.
However, the modularity of the design of \FfF step allows for much easier scaling.
The base flow can be trained on the entire patch and frozen and then individual top flows can be trained on respective regions of interest, for efficient scaling.
\skyc builds the template by leveraging the correlations of features with the proper motions, which results in a more background representative template.
This allows the method to use more discriminatory features to be used in the downstream task regardless of their correlation with the proper motions, which otherwise would lead to a biased classifier and thus false positives.

The follow-up work will involve applying \skyc in a full sky search for stellar streams in the Milky Way, and comparing the performance of the method with other existing methods.
The latest data release GDR3 from \gaia contains over 1.8 billion sources with improved astrometric and photometric measurements for a significantly larger number of sources compared to GDR2, which will provide a more detailed view of the Milky Way.
The improved measurements of radial velocities of sources in GDR3 will also allow for a more detailed study of the kinematics of the streams.
The improvements in data quality in GDR3 would likely further improve the performance of \skyc and as such is a natural next step for the method.

\section*{Acknowledgements}
We would like to acknowledge funding through the SNSF Sinergia grant CRSII5\_193716 ``Robust Deep Density Models for High-Energy Particle Physics and Solar Flare Analysis (RODEM)''.
This work has made use of data from the European Space Agency (ESA) mission {\gaia} (\url{https://www.cosmos.esa.int/gaia}), processed by the {\gaia} Data Processing and Analysis Consortium (DPAC, \url{https://www.cosmos.esa.int/web/gaia/dpac/consortium}).
Funding for the DPAC has been provided by national institutions, in particular the institutions participating in the {\it Gaia} Multilateral Agreement.
The computations were performed at University of Geneva using Baobab HPC service.
Special thanks to Samuel Klein and Kinga Anna Wozniak for their inputs during the development of the method and this manuscript.

\section*{Data Availability}

\skyc uses the publicly available GDR2 data.
A curated set of 21 patches used in this work is available at \url{https://zenodo.org/records/7897936}.
The \gdone membership labels are taken from \url{https://zenodo.org/records/1295543}.

\bibliographystyle{mnras}
\bibliography{main.bib}

\appendix
\section{\FfF Training and Hyperparameter Tuning Details}
\label{sec:appendix_tuning}

\subsection{\FfF features preprocessing}
The first step in training the \FfF model is to define the SR and SB regions in \pmlat.
In this work, we chose the SR in a given patch to ensure that the \gdone is fully contained.
The SB region, defined as the region adjacent to the SR is chosen to be $\sim 6$ mas/yr wide.
This ensures sufficient training statistics for the base and the top flow model.
The features used for training the \FfF model are: [\lon, \lat, \magnitude, \gbprp, \pmlonstar] and the conditional feature is \pmlat.
As these features have different dynamic ranges, we opt to further scale them to ensure a stable model training.
All features are first scaled to be in the range $[0, 1]$.
The \magnitude feature has a sharp cutoff at 20.2 which proves to be a difficult feature for generative models to learn.
To mitigate this, we apply a logit transformation to the \magnitude feature.
The logit transformation is defined as:
\begin{equation}
    \text{logit}(x) = \log\left(\frac{x}{1-x}\right),
\end{equation}
Finally, all features are scaled to be in the range $[-3, 3]$.

The data for the base and top flow training is divided into training and validation sets in a $80:20$ ratio.
\subsection{Hyperparameter tuning}
The three neural network components in the \skyc method are the base flow, the top flow, and the classifier in the \cwola step.
The base flow comprises of a stack of autoregressive transformations parametrised by a Rational Quadratic Spline (RQS) function.
The top flow is a stack of coupling transformations, also parametrised by a RQS function.
The hyperparameters for the base and the top flow are the number of stacked transformations, the number of bins in the RQS function, and the number of hidden units and layers in the multi-layer perceptron (MLP) used to estimate the parameters of the RQS function.
Other hyperparameters include the learning rate, the batch size, and the number of epochs for training the base and top flow.
The classifier in the \cwola step is an MLP with 3 hidden layers of 32 units, and are trained using the Adam optimizer with a learning rate of $10^{-3}$ with k-Fold cross validation (with k = 5).
We found this architecture to be robust across different patches, and did not perform any hyperparameter scans.
For the \FfF training, we want to ensure that the generated template is in accordance with data.
Since we do not know a priori if there is a signal in the SR, we test the performance of the \FfF model in the sidebands.
We select the hyperparameters that minimise the AUC score for the \cwola classifier on \SBone and \SBtwo vs template classification.
For a well-trained \FfF model the generated template should have an AUC score close to 0.5.

The hyperparameters for the base and top flow are listed in Table~\ref{tab:hyperparameters}, where the hyperparameters for the base flows were found to give robust performance regardless of the patch, and so held constant.
For the top flows, there could be significant variation in performance related to the hyperparameter selection depending on the patch, and so hyperparameter tuning was performed to find the values that performed well regardless of the patch.
Both base and top flow were capped at a maximum number of 150, and 100 epochs respectively.
While the base flow seemed to improve with higher number of epochs, the top flow converged much more quickly at $\sim 30-40$ epochs of training.
\begin{table}
    \centering
    \caption{Hyperparameters for \FfF Training}
    \label{tab:hyperparameters}
    \begin{tabular}{clr}
        \hline
        \textbf{Model} & \textbf{Hyperparameter Name} & \textbf{Value} \\
        \hline \hline
        \multirow{7}{*}{Base Flow} & Number of Stacked Transformations & 4 \\
        & Number of Bins in RQS Function & 12 \\
        & Number of Hidden Units in MLP & 64 \\
        & Number of Layers in MLP & 2 \\
        & Learning Rate & 0.0001 \\
        & Batch Size & 512 \\
        & Maximum number of Epochs & 150 \\
        \hline
        \multirow{7}{*}{Top Flow} & Number of Stacked Transformations & 6 \\
        & Number of Bins in RQS Function & 10 \\
        & Number of Hidden Units in MLP & 32 \\
        & Number of Layers in MLP & 2 \\
        & Learning Rate & 0.001 \\
        & Batch Size & 512 \\
        & Maximum number of Epochs & 100 \\
        \hline
    \end{tabular}
\end{table}

\end{document}